\documentclass[aps,prb,reprint,longbibliography, superscriptaddresses]{revtex4-2}
\usepackage{mathrsfs}
\usepackage{textcomp, gensymb}
\usepackage{amsmath,gensymb}
\usepackage{amsfonts}
\usepackage{amssymb}
\usepackage{amsthm}
\usepackage{graphicx}
\usepackage{natbib}
\usepackage{xcolor}
\usepackage{hyperref}
\usepackage{bm}
\usepackage[caption=false]{subfig}
\usepackage{verbatim}
\usepackage{braket}
\usepackage{physics}
\usepackage{ulem}

\newcommand{\alr}[1]{{\color{black} #1}}
\newcommand{\ak}[1]{{\color{black} #1}}

\begin{document}
	\title{Quantum Sensing of Antiferromagnetic Magnon Two-Mode Squeezed Vacuum}
	
	\author{Anna-Luisa E. R\"omling} 
	\email{anna-luisa.romling@uam.es}
	\affiliation{Condensed Matter Physics Center (IFIMAC) and Departamento de F\'isica
		Te\'orica de la Materia Condensada, Universidad Aut\'onoma de Madrid,
		E-28049 Madrid, Spain}
	
	\author{Akashdeep Kamra}
	\affiliation{Condensed Matter Physics Center (IFIMAC) and Departamento de F\'isica
		Te\'orica de la Materia Condensada, Universidad Aut\'onoma de Madrid,
		E-28049 Madrid, Spain}
	
	\begin{abstract}
		N\'eel ordered antiferromagnets exhibit two-mode squeezing such that their ground state is a nonclassical superposition of magnon Fock states. Here we theoretically demonstrate that antiferromagnets can couple to spin qubits via direct dispersive interaction stemming from, e.g., interfacial exchange. We demonstrate that this kind of coupling induces a magnon number dependent level splitting of the excited state resulting in multiple system excitation energies. This series of level splittings manifests itself as nontrivial excitation peaks in qubit spectroscopy thereby revealing the underlying nonclassical magnon composition of the antiferromagnetic quantum state. By appropriately choosing the drive or excitation energy, the magnonic state can be controlled via the qubit, suggesting that Fock states of magnon pairs can be generated deterministically. This enables achieving states useful for quantum computing and quantum information science protocols.
	\end{abstract}
	\maketitle

	\section{Introduction \label{sec:intro}}
	Antiferromagnets (AFMs) are materials with magnetic order and a vanishing net macroscopic magnetization~\cite{rezende2019}. Due to their robustness against magnetic fields, their fast THz dynamics and phenomena such as exchange bias and spin-orbit effects, AFMs have been investigated especially for their potential in spintronics~\cite{hoffmann2022, fukami2020, han2023,baltz2018,yan2020,zelezny2018}. The classical AFM ground state can be described by a N\'eel ordered state comprising two sublattices of oppositely oriented spins~\cite{mattis1963}. Coherent excitations on the magnetic order generate a collective precession of the magnetic moments around their equilibrium position referred to as spin waves or magnons~\cite{violakusminskiy2019,han2023}. Being two-sublattice magnets, easy-axis collinear AFMs host two kinds of spin waves that are distinguished by chirality~\cite{han2023}. While the semi-classical spin wave description is successful in explaining many phenomena~\cite{han2023}, it misses important physics~\cite{kamra2019,hirjibehedin2006,wuhrer2022} as the true quantum ground state of the ordered AFM is superposition of states with an equal number of spin-up and spin-down magnons~\cite{kamra2019}. Therefore, the AFM ground state is \ak{nonclassical and harbors composite excitations capable of generating states useful for quantum information protocols~\cite{ladd2010,yuan2022,pirro2021,barman2021,skogvoll2021,kamra2019}.} \alr{It is therefore important to establish protocols to detect and quantify the quantum properties of these states.} 
	
	If two observables are non-commuting, their quantum fluctuations obey Heisenberg's uncertainty principle~\cite{heisenberg1927}. Squeezing is a phenomenon where the quantum fluctuations of one observable are reduced beyond the standard quantum limit at the expense of the other's~\cite{walls1983, walls2008}. In quantum optics, squeezed states of light have been exploited in feats such as the detection of gravitational waves due to reduced quantum noise~\cite{grote2013}. While squeezed states of light are nonquilibrium, magnets exhibit squeezing in equilibrium suggesting to be a useful platform for quantum computing purposes~\cite{kamra2016,kamra2019,wuhrer2022,tabuchi2016,xu2023a}. If a ferromagnet possesses anisotropy, the ground state fluctuation of the total spin $S_x$ and $S_y$ components are squeezed, adjusting to the energy cost dictated by the magnet's anisotropy. \ak{Considering the spatially homogeneous component corresponding to wavevector $\boldsymbol{k} = \boldsymbol{0}$, the} ferromagnetic ground state becomes a one-mode squeezed-magnon vacuum which is composed of even magnon number states making the ground state a nonclassical superposition~\cite{kamra2016,walls2008}. While ferromagnets exhibit squeezing only in the presence of anisotropies, AFMs display (typically large) squeezing as an intrinsic property due to strong exchange interaction between the sublattices~\cite{yuan2020,kamra2019}. As a consequence, the fluctuations of the total spin of the two sublattices become quantum correlated, such that the AFM exhibits two-mode squeezing of the two sublattice modes. As a result, the ground state is a two-mode squeezed vacuum -- a superposition of entangled pairs of spin-up and spin-down sublattice modes~\cite{kamra2019}. \ak{Ferromagnets also exhibit two-mode squeezing between the $+\boldsymbol{k}$ and $-\boldsymbol{k}$ magnon modes~\cite{kamra2016}.} Two-mode squeezed states exhibit Bell nonlocality~\cite{brunner2014,azimimousolou2021}, making them useful for manipulation of quantum information and study of fundamental principles of quantum mechanics~\cite{pirandola2015,brunner2014,rossatto2011}. While there have been experiments successfully realizing two-mode spin squeezing with coherent drives~\cite{walls2008,kamra2019,zhao2004,zhao2006,bossini2019}, generating squeezed magnon Fock states poses a bigger challenge. This, in part, motivates the present study to generate two-mode squeezed states in AFMs and the design of experiments to study their quantum properties.  
	
	 \ak{The} detection of AFM structure is challenging due to insensitivity to magnetic fields and makes investigating the quantum nature of AFMs an experimental challenge~\cite{jungwirth2016,baltz2018}. A recently suggested theoretical protocol~\cite{azimi-mousolou2023} finds that the two-mode magnon entanglement exhibited by AFMs can be detected via modulation of Rabi oscillations in a transmon qubit. While the number state resolution of nonequilibrium states has been demonstrated experimentally for photons~\cite{schuster2007,boissonneault2009,kono2017} and magnons~\cite{lachance-quirion2017}, a recent theoretical proposal suggests that the equilibrium magnon composition of a ferromagnetic ground state~\cite{kamra2016} can also be resolved via a qubit~\cite{romling2023}. This proposal is based on a unique direct dispersive magnon-qubit coupling stemming from, e.g., interfacial exchange interaction~\cite{kamra2017a}. The coupling induces a magnon number dependent excitation energy of the qubit, enabling to generate magnon number states via controlled drive of the qubit and revealing the ground state composition related to squeezing via qubit spectroscopy. Since AFMs potentially also offer interfacial exchange interaction~\cite{kamra2018}, we are motivated to examine control and probe of the nonclassical magnon states in an AFM via a spin qubit. \ak{This qubit-based approach is complementary to the recent suggestions~\cite{lee2023a,wuhrer2023} of employing light for probing the squeezed nature of magnetic states and excitations.} 
	
\begin{figure}[tb]
		\includegraphics[width=\columnwidth]{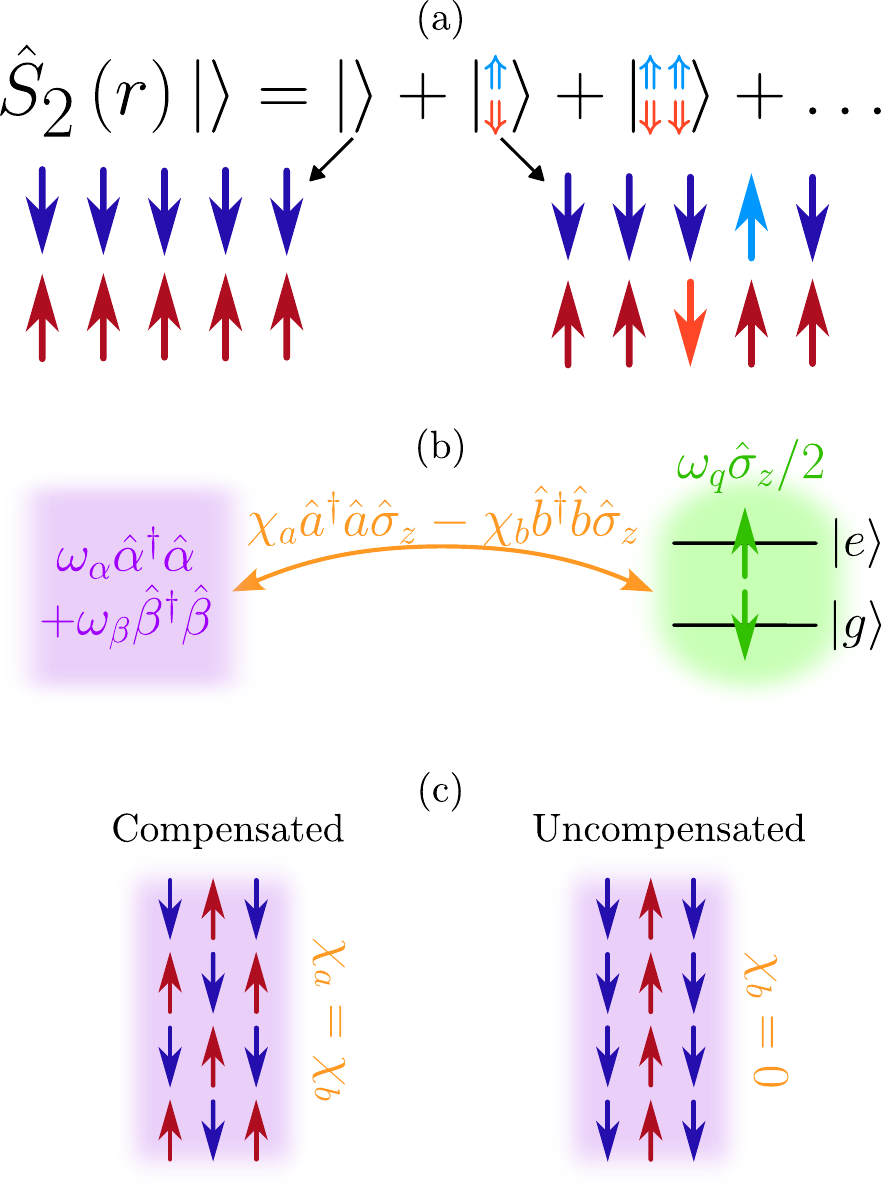}
		\caption{Schematic depiction of (a) two-mode squeezed magnon vacuum (b) magnon-qubit interaction and (c) interfaces of the antiferromagnet. (a) The two-mode squeeze operator $\hat{S}_2(r)$ applied to the N\'eel ordered state $\ket{}$ is a superposition of states with equal number of delocalized spin flips on the down-spin (blue) and up-spin (red) sublattices~\cite{kamra2019}. (b) The spin qubit $\hat{\sigma}_z$ (green) couples to the antiferromagnetic sublattice (spin flip) magnons $\hat{a}$ and $\hat{b}$ via direct dispersive interaction~\cite{romling2023} $\chi_a \hat{a}^\dagger \hat{a} - \chi_b \hat{b}^\dagger \hat{b} $ (orange). The eigenmodes of the antiferromagnet are two-mode squeezed-magnons $\hat{\alpha}$ and $\hat{\beta}$ (purple). (c) A compensated interface consists of equal number of up and down spins corresponding to equal direct dispersive coupling strength $\chi_a=\chi_b$ (left). In a completely uncompensated interface, $\chi_a$ is maximized while $\chi_b=0$ (right)~\cite{kamra2017a}.\label{fig:Fig1}}
	\end{figure}

	Here, we investigate a two-sublattice AFM. Its quantum ground state, a two-mode squeezed magnon vacuum, is a superposition of states with an equal number of spin-up and spin-down magnons [Fig.~\ref{fig:Fig1}(a)]~\cite{kamra2019}. We theoretically demonstrate that the AFM can be coupled to a spin qubit via direct dispersive interaction [Fig.~\ref{fig:Fig1}(b)]~\cite{romling2023} and that the ground state of the coupled system is a superposition of excited states with an equal number of spin-up and spin-down squeezed-magnons. These levels are non-degenerate such that there are multiple magnon-dependent qubit excitation energies, allowing to control the magnonic state and probe the magnon composition of the AFM ground state via the qubit. While the large frequency of AFM magnons makes the probe via a qubit difficult, we find that excitation probabilities under the qubit drive are exponentially enhanced via the squeezing present in the AFM~\cite{qin2018,leroux2018,lu2015,skogvoll2021,yuan2020}. We expect this squeezing-mediated enhancement to be strong due to the large squeezing present in AFMs and therefore probing the AFM ground state easier to achieve. We show that the direct dispersive coupling stems from interfacial spin exchange interaction~\cite{kamra2017a} and that the structure of the AFM interface determines the coupling strength, i.e. if the interface is compensated or uncompensated [Fig.~\ref{fig:Fig1}(c)]. Our findings suggest that the qubit probe and control only shows nontrivial effects if the AFM interface is uncompensated. 
	
	We structure the paper as follows: In Sec.~\ref{sec:boson_model} we develop and analyze the bosonic model for an AFM coupled to a spin qubit. In Sec.~\ref{sec:physical_realizations} we turn our attention to realistic systems and derive the bosonic Hamiltonian for hematite and the direct dispersive coupling from a spin model. Finally, we discuss and conclude our paper in Secs.~\ref{sec:discussion} and \ref{sec:conclusions}.  
	
	\section{Antiferromagnetic magnons dispersively coupled to a qubit \label{sec:boson_model}}
	In this section, we analyse the quantum model of a two-sublattice AFM coupled to a spin qubit via direct dispersive interaction~\cite{romling2023}. We discuss the uncoupled AFM and introduce useful notation and functions. We then determine the eigenmodes and eigenenergies of the coupled system via projection onto the qubit ground ($\ket{g}$) and excited ($\ket{e}$) states. The resulting reduced Hamiltonians reveal the system ground state and a set of excited states. Finally, we discuss probe and control of the system state via controlled drive of the qubit $\ket{g} \rightarrow \ket{e}$.  
	\subsection{Uncoupled antiferromagnet \label{sec:uncoupled_afm}}
	
	We consider an AFM with Heisenberg exchange and Dzyaloshinskii–Moriya interaction between neighboring spins $\hat{\boldsymbol{S}}_i$ and $\hat{\boldsymbol{S}}_j$, easy-axis anisotropy in $\hat{\boldsymbol{z}}$-direction and an applied magnetic field along the easy-axis. In Sec.~\ref{sec:hematite}, we present the corresponding full spin Hamiltonian. Here, starting from N\'eel ordering, we define two sublattices: Sublattice A is the sublattice of all spins pointing along $-\hat{\boldsymbol{z}}$ and sublattice $B$ is the sublattice of all spins pointing along $+\hat{\boldsymbol{z}}$~\cite{kamra2019,kamra2017,kittel1987} [see Fig.~\ref{fig:Fig1}(a)]. We map spin operators $\hat{\boldsymbol{S}}_i$ and $\hat{\boldsymbol{S}}_j$ onto boson operators $\hat{a}_i$ and $\hat{b}_j$ via  Holstein-Primakoff transformations~\cite{holstein1940} and switch to Fourier space [see Eqs.~\eqref{eq:HP1}--\eqref{eq:HP6} in Sec.~\ref{sec:hematite} for details]. Considering a small magnet, we only take the uniform $\boldsymbol{k} = \boldsymbol{0}$ mode into account~\cite{skogvoll2021}. We define bosonic excitations on the two sublattices via $\hat{S}_{i\in A}^+\propto\hat{a}_{\boldsymbol{k}=\boldsymbol{0}}^\dagger$ and $\hat{S}_{j\in B}^-\propto\hat{b}_{\boldsymbol{k}=\boldsymbol{0}}^\dagger$ interpreting them as delocalized spin flips on sublattice A and B respectively [see Fig.~\ref{fig:Fig1}(a)]. We drop the index $\boldsymbol{k}=\boldsymbol{0}$ and refer to them as spin-up ($\hat{a}$) and spin-down ($\hat{b}$) sublattice-magnons. We obtain the following Hamiltonian~\cite{kamra2019,kamra2017,kittel1987} 
	\begin{align}
		\hat{\mathcal{H}}_{\mathrm{AFM}}=	A\hat{a}^{\dagger}\hat{a}+&B\hat{b}^{\dagger}\hat{b}+C^*\hat{a}\hat{b}+C\hat{a}^{\dagger}\hat{b}^{\dagger}, \label{eq:H_afm}
	\end{align}
	where the parameters $A$ and $B$ are evaluated to be real from the spin model. The parameter $C$ can be complex and quantifies the exchange between the two sublattices [see Sec.~\ref{sec:hematite} for the full derivation]. Our goal is to diagonalize the Hamiltonian $\hat{\mathcal{H}}_\text{AFM}$ [Eq.~\eqref{eq:H_afm}] to understand the eigenmodes of the AFM and its quantum ground state. 	
	
	In the following, it will be convenient to define the functions 
	\begin{align}
		\mathcal{E}\left(x,y\right)&=\sqrt{\left(x+y\right)^{2}/4-\left|C\right|^{2}} \label{eq:E} \\ 
		\Delta\left(x,y\right) &= \left(x-y\right)/2 \label{eq:Delta},\\ 
		\mathcal{U}\left(x,y\right)&=\frac{1}{\sqrt{2}}\sqrt{\frac{x+y}{2\mathcal{E}\left(x,y\right)}+1}, \label{eq:U}\\
		\mathcal{V}\left(x,y\right)&=\frac{e^{i\phi}}{\sqrt{2}}\sqrt{\frac{x+y}{2\mathcal{E}\left(x,y\right)}-1}, \label{eq:V}
	\end{align}
	with the phase $e^{i\phi}=C/\left|C\right|$. We diagonalize the Hamiltonian $\hat{\mathcal{H}}_\mathrm{AFM}$ [Eq.~\eqref{eq:H_afm}] via the Bogoliubov transformations~\cite{bogoljubov1958, kittel1987} $\hat{\alpha}=u\hat{a}+v\hat{b}^{\dagger}$ and $\hat{\beta}=u\hat{b}+v\hat{a}^{\dagger}$, where the Bogoliubov coefficients are given by $u = \mathcal{U}\left(A,B\right)$ and $v = \mathcal{V}\left(A,B\right)$ [Eqs.~\eqref{eq:U} and \eqref{eq:V}]. We obtain the diagonalized Hamiltonian $\hat{\mathcal{H}}_\text{AFM} = \omega_\alpha \hat{\alpha}^\dagger\hat{\alpha} + \omega_\beta \hat{\beta}^\dagger\hat{\beta} + E_0$ with the eigenfrequencies $\omega_{\alpha/\beta}$ and a zero-point energy $E_0 = \left(\omega_\alpha + \omega_\beta - A -B \right)/2$. The eigenfrequencies explicitly read
	\begin{align}
		\omega_{\alpha}	&=\mathcal{E}\left(A,B\right)+\Delta\left(A,B\right), \label{eq:omega_alpha}\\
		\omega_{\beta}	&=\mathcal{E}\left(A,B\right) -\Delta\left(A,B\right) \label{eq:omega_beta},
	\end{align}
	where we used the functions $\mathcal{E}\left(x,y\right)$ and $\Delta\left(x,y\right)$ [Eqs.~\eqref{eq:E} and \eqref{eq:Delta}]. It is convenient to denote the average of the eigenfrequencies as $\varepsilon = \left(\omega_\alpha + \omega_\beta\right)/2$ which corresponds to the AFM resonance without an applied magnetic field.
	
	The eigenmodes ($\hat{\alpha}$ and $\hat{\beta}$) are related to the sublattice-magnons ($\hat{a}$ and $\hat{b}$) via the two-mode squeeze transformations $\hat{\alpha} = \hat{S}_{2}\!\left(\xi\right) \hat{a} \hat{S}_{2}^\dagger\left(\xi\right)$ and $\hat{\beta} = \hat{S}_{2}\!\left(\xi\right) \hat{b} \hat{S}_{2}^\dagger\left(\xi\right)$~\cite{kamra2019,gerry2005,walls2008}. The two-mode squeeze operator $\hat{S}_{2}\!\left(\xi\right)$ is defined by~\cite{caves1985, walls2008,gerry2005} 
	\begin{equation}
		\hat{S}_{2}\!\left(\xi\right)	=\exp(\xi^*\hat{a}\hat{b}-\xi\hat{a}^{\dagger}\hat{b}^{\dagger}), \label{eq:S2}
	\end{equation} 
	with the two-mode squeeze factor $\xi = r e^{i\phi}$. Its absolute value is given by
	\begin{equation}
		\tanh(r)=\left|\mathcal{V}\left(A,B\right)\right|/\mathcal{U}\left(A,B\right), \label{eq:r} 
	\end{equation} 
	and the phase $\phi$ is the same as in Eq.~\eqref{eq:V}. Note that per our definition, $\mathcal{U}\left(A,B\right)$ [Eq.~\eqref{eq:U}] is real and positive, such that $r$ [Eq.~\eqref{eq:r}] is real and positive as well. We refer to the AFM eigenexcitations as spin-up ($\hat{\alpha}$) and spin-down ($\hat{\beta}$) two-mode squeezed-magnons (TMSMs).
	
\begin{figure*}[tb]
		\onecolumngrid
		\includegraphics[width=\textwidth]{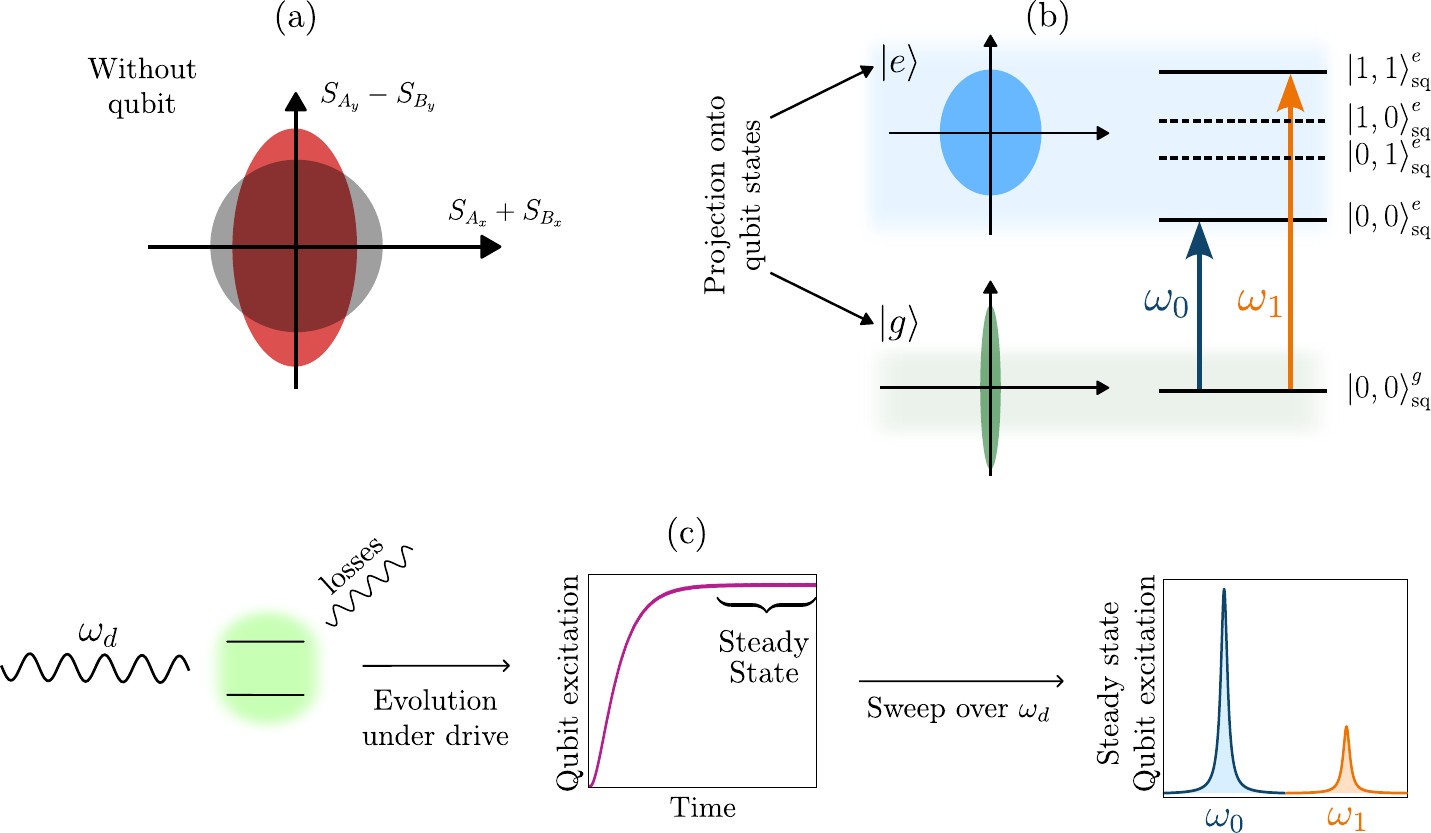}
		\caption{Schemetic depiction of (a) quantum fluctuations of the two-mode squeezed magnon vacuum (b) energy levels of the ground state and the first 4 excited states and (c) qubit spectroscopy protocol. (a) We depict the phase space of $S_{A_x}+S_{B_x}$ and $S_{A_y}-S_{B_y}$, corresponding to to the sum of $x$ -and difference of $y$-components of the total spin on sublattice A and B and compare the quantum fluctuation of the two-mode squeezed magnon vacuum $\ket{0,0}_\mathrm{sq} = \hat{S}_2\ket{0,0}_\mathrm{sub}$ with an elliptical shape (red) to the isotropic quantum fluctuations of the sublattice-magnon vacuum $\ket{0,0}_\mathrm{sub}$ (grey)~\cite{kamra2020}. (b) We depict the two subspaces resulting from projection onto the qubit states $\ket{g}$ and $\ket{e}$, with squeezing $r_g$ (green) and $r_e$ (blue). The ground state $\ket{0,0}_\mathrm{sq}^g$ can be excited into states with an equal number of spin-up and spin-down two-mode squeezed magnons $\ket{n,n}_\mathrm{sq}^e$ by driving the qubit with frequency $\omega_d=\omega_n$ [Eq.~\eqref{eq:omega_n_full}]. (c) We depict a lossy qubit that is driven by a monochromatic microwave drive with frequency $\omega_d$. The qubit evolves under the drive until a steady state is reached. Sweeping over drive frequency $\omega_d$ and measuring steady state qubit excitation for each $\omega_d$ results in the qubit spectrum with excitation peaks at frequencies $\omega_0$, $\omega_1,\dots$ [Eq.~\eqref{eq:omega_n_full}]. \label{fig:Fig2}}
	\end{figure*}

	We denote the eigenstates of $\hat{\mathcal{H}}_\text{AFM}$ [Eq.~\eqref{eq:H_afm}] by $\ket{n_\alpha,n_\beta}_\mathrm{sq}$ which is a TMSM Fock state with $n_\alpha$ spin-up and $n_\beta$ spin-down TMSM excitations. The AFM ground state is the TMSM vacuum denoted by $\ket{0, 0}_\mathrm{sq}$~\cite{kamra2019}. The TMSM vacuum can be obtained from the sublattice-magnon vacuum $\ket{0, 0}_\mathrm{sub}$ (absence of spin flips) via $\ket{0, 0}_\mathrm{sq} = \hat{S}_{2}\!\left(\xi\right) \ket{0, 0}_\mathrm{sub}$. Denoting $\ket{n_a,n_b}_\mathrm{sub}$ as a Fock state with $n_a$ spin-up and $n_b$ spin-down sublattice-magnons, our TMSM vacuum can be expanded as~\cite{kamra2019,gerry2005,walls2008}
	\begin{equation}
		\ket{0,0}_{\mathrm{sq}}	=\sum_{n} \frac{\left[-e^{i\phi}\tanh(r)\right]^{n}}{\cosh(r)} \ket{n,n}_\mathrm{sub}. \label{eq:gs_sp}
	\end{equation}
	The AFM ground state is therefore a superposition of states with equal number of delocalized spin flips on both sublattices [see Fig.~\ref{fig:Fig1}(a)]. Due to Heisenberg uncertainty relation~\cite{cohen-tannoudji1977}, the total spin on sublattice A and B fluctuate. Because of the strong exchange interaction between the sublattices, these fluctuations are quantum correlated such that the ground state exhibits squeezed quantum fluctuations~\cite{kamra2020, kamra2019} [see Fig.~\ref{fig:Fig2}(a)].
	
	\subsection{Eigenmodes of the coupled system \label{sec:coupled_system}}
	
	We consider an AFM as discussed in Sec.~\ref{sec:uncoupled_afm}, described by Hamiltonian $\hat{\mathcal{H}}_\text{AFM}$ [Eq.~\eqref{eq:H_afm}], and couple it to a spin qubit $\hat{\sigma}_z$. As depicted in Fig.~\ref{fig:Fig1}(b) and (c), the qubit couples to the spins in the antiferromagnetic interface via spin exchange interaction $\propto \hat{\boldsymbol{S}}\cdot \hat{\boldsymbol{\sigma}}$. As detailed in Sec.~\ref{sec:interface}, the dominating contribution comes from the $z$-component of the interfacial exchange if the bare AFM eigenfrequencies $\omega_\alpha$ and $\omega_\beta$ [Eqs.~\eqref{eq:omega_alpha} and \eqref{eq:omega_beta}] are far detuned from the bare qubit level splitting $\omega_q$. From this, we obtain a direct dispersive coupling between the spin qubit and sublattice-magnons $\hat{a}$ and $\hat{b}$ and find that the coupling strength depends on the size of the magnet and the structure of the interface. Assuming a large detuning between AFM and spin qubit, we neglect coherent exchange~\cite{gerry2005, skogvoll2021}. As demonstrated in Sec.~\ref{sec:interface}, we obtain the following Hamiltonian describing the coupled magnet-qubit system  
	\begin{align}
		\hat{\mathcal{H}}_{0}=	A\hat{a}^{\dagger}\hat{a}+&B\hat{b}^{\dagger}\hat{b}+C^*\hat{a}\hat{b}+C\hat{a}^{\dagger}\hat{b}^{\dagger}+\frac{\omega_{q}}{2}\hat{\sigma}_{z} \nonumber \\  &+\chi_{a}\hat{a}^{\dagger}\hat{a}\hat{\sigma}_{z}-\chi_{b}\hat{b}^{\dagger}\hat{b}\hat{\sigma}_{z}, \label{eq:H_0} 
	\end{align}
	where the parameters $\chi_{a}$ and $\chi_{b}$ quantify the direct dispersive coupling strength and are assumed to be real and positive for simplicity. 
	
	We start to diagonalize the Hamiltonian $\hat{\mathcal{H}}_0$ [Eq.~\eqref{eq:H_0}] by projecting onto the qubit ground ($\ket{g}$) and excited ($\ket{e}$) states. We find that the reduced Hamiltonian $\hat{\mathcal{H}}_g = \bra{g}\hat{\mathcal{H}}_0\ket{g}$ is given by
	\begin{align}
		\hat{\mathcal{H}}_{g}	=&\left(A-\chi_{a}\right) \hat{a}^{\dagger}\hat{a}+\left(B+\chi_{b}\right)\hat{b}^{\dagger}\hat{b} \nonumber\\
		&+C^*\hat{a}\hat{b}+C\hat{a}^{\dagger}\hat{b}^{\dagger}-\frac{\omega_{q}}{2}. \label{eq:H_g}
	\end{align}
	The excited state projection $\hat{\mathcal{H}}_e = \bra{e}\hat{\mathcal{H}}_0\ket{e}$ can be obtained from the expression for $\hat{\mathcal{H}}_g$ [Eq.~\eqref{eq:H_g}] \alr{upon substitutions $-\chi_a \rightarrow + \chi_a$, $+\chi_b \rightarrow - \chi_b$ and $-\omega_q \rightarrow + \omega_q$}. 
	
	We diagonalize the reduced Hamiltonians $\hat{\mathcal{H}}_{g}$ and $\hat{\mathcal{H}}_{e}$ [Eq.~\eqref{eq:H_g}] via the following Bogoliubov transformations~\cite{bogoljubov1958}, 
	\begin{align}
		\hat{\alpha}_{g/e}	&=u_{g/e}\hat{a}+v_{g/e}\hat{b}^{\dagger}, \label{eq:alpha_g_e}\\ 
		\hat{\beta}_{g/e}	&=u_{g/e}\hat{b}+v_{g/e}\hat{a}^{\dagger}, \label{eq:beta_g_e}
	\end{align}
	with Bogoliubov coefficients 
	\begin{align}
		u_{g}	&=\mathcal{U}\left(A- \chi_a,B+\chi_b\right), \label{eq:u_g}\\
		v_{g}	&=\mathcal{V}\left(A- \chi_a,B+\chi_b\right), \label{eq:v_g}\\
		u_{e}	&=\mathcal{U}\left(A+ \chi_a,B-\chi_b\right), \label{eq:u_e}\\
		v_{e}	&=\mathcal{V}\left(A+ \chi_a,B-\chi_b\right). \label{eq:v_e}
	\end{align}
	Note that we use the labels $g$ and $e$ to denote the state of the qubit. Applying the Bogoliubov transformations [Eqs.~\eqref{eq:alpha_g_e} and \eqref{eq:beta_g_e}], we obtain the following diagonalized Hamiltonians
	\begin{align}
		\hat{\mathcal{H}}_{g} = &\omega_\alpha^g \hat{\alpha}_{g}^\dagger\hat{\alpha}_{g} + \omega_\beta^g \hat{\beta}_{g}^\dagger\hat{\beta}_{g} \nonumber\\&+ \frac{1}{2} \left(\omega_\alpha^g + \omega_\beta^g - A - B + \chi_a - \chi_b -\omega_q \right), \label{eq:H_g_diag}\\
		\hat{\mathcal{H}}_{e} = &\omega_\alpha^e \hat{\alpha}_{e}^\dagger\hat{\alpha}_{e} + \omega_\beta^e \hat{\beta}_{e}^\dagger\hat{\beta}_{e} \nonumber\\&+ \frac{1}{2} \left(\omega_\alpha^e + \omega_\beta^e - A - B - \chi_a + \chi_b + \omega_q \right), \label{eq:H_e}
	\end{align}
	with eigenfrequencies 
	\begin{align}
		\omega_{\alpha}^{g}	&=\mathcal{E}\left(A- \chi_a,B+\chi_b\right) +\Delta\left(A- \chi_a,B+\chi_b\right), \label{eq:omega_alpha_g}\\ 
		\omega_{\beta}^{g}	&=\mathcal{E}\left(A- \chi_a,B+\chi_b\right) -\Delta\left(A- \chi_a,B+\chi_b\right), \label{eq:omega_beta_g}\\
		\omega_{\alpha}^{e}	&=\mathcal{E}\left(A+ \chi_a,B-\chi_b\right) +\Delta\left(A+ \chi_a,B-\chi_b\right), \label{eq:omega_alpha_e}\\
		\omega_{\beta}^{e}	&=\mathcal{E}\left(A+ \chi_a,B-\chi_b\right) -\Delta\left(A+ \chi_a,B-\chi_b\right) \label{eq:omega_beta_e}.
	\end{align}
	For later, it will be convenient to define the average frequencies $\varepsilon_{g/e}= \left(\omega_{\alpha}^{g/e}+\omega_{\beta}^{g/e}\right)/2$. 
	
	Similar to $\hat{\alpha}$ and $\hat{\beta}$ discussed in Sec.~\ref{sec:uncoupled_afm}, the eigenmodes $\hat{\alpha}_{g/e}$ and $\hat{\beta}_{g/e}$ [Eqs.~\eqref{eq:alpha_g_e} and \eqref{eq:beta_g_e}] are spin-up and spin-down TMSM. We obtain that their squeeze factors are given by $\xi_{g/e} = r_{g/e} e^{i\phi}$, having the same phase $\phi$ as $\xi$ for the uncoupled AFM in Sec.~\ref{sec:uncoupled_afm}. We find that their respective absolute values $r_{g/e}$ are given by the relations
	\begin{equation}
		\tanh(r_{g/e})=\frac{\left|v_{g/e}\right|}{u_{g/e}}, \label{eq:r_g_e}
	\end{equation}
	where we used Bogoliubov coefficients [Eqs.~\eqref{eq:u_g}--\eqref{eq:v_e}]. All of the Bogoliubov coefficients $u_{g/e}$ and $v_{g/e}$ [Eqs.~\eqref{eq:u_g}--\eqref{eq:v_e}] depend explicitly on the difference between the direct dispersive coupling strengths $(\chi_a - \chi_b)$. If the direct dispersive coupling strengths are the same $\chi_a = \chi_b$ then $u_g=u_e = u$ and $v_v=v_e = v$. This follows from the invariances $\mathcal{U}\left(x+\lambda,y-\lambda\right) = \mathcal{U}\left(x,y\right)$ and $\mathcal{V}\left(x+\lambda,y-\lambda\right) = \mathcal{V}\left(x,y\right)$ [Eqs.~\eqref{eq:U} and \eqref{eq:V}]. Inserting the relations $(A+B)=(\omega_\alpha+\omega_\beta)\cosh(2r)$ and $
	\left|C\right| = (\omega_\alpha+\omega_\beta)\left|\sinh(2r)\right|/2$ [obtained from Eqs.~\eqref{eq:omega_alpha}, \eqref{eq:omega_beta} and \eqref{eq:r}], one can show that $u_{g/e}$ and $\left|v_{g/e}\right|$ [Eqs.~\eqref{eq:u_g}--\eqref{eq:v_e}] only depend on squeezing $r$ and the ratio $\left(\chi_a-\chi_b\right)/\left(\omega_\alpha + \omega_\beta\right)$. Therefore, $r_g$ and $r_e$ [Eq.~\eqref{eq:r_g_e}] are functions of $r$ and $\left(\chi_a-\chi_b\right)/\left(\omega_\alpha + \omega_\beta\right)$ and have different values if the direct dispersive coupling strengths fulfil $\chi_a \neq \chi_b$. 
	
	We denote $\hat{\alpha}_g$ and $\hat{\beta}_g$ as ground state TMSM with squeeze factor $\xi_g$ and correspondingly the excited state TMSM $\hat{\alpha}_e$ and $\hat{\beta}_e$ with $\xi_e$. We conclude that the eigenstates of $\hat{\mathcal{H}}_g$ and $\hat{\mathcal{H}}_e$ [Eqs.~\eqref{eq:H_g_diag} and \eqref{eq:H_e}] are given by TMSM Fock states. We denote them by $\ket{n_\alpha^g, n_\beta^g}_\mathrm{sq}^g$ and $\ket{n_\alpha^e, n_\beta^e}_\mathrm{sq}^e$, where $n_{\alpha(\beta)}^g$ is the number of spin-up (spin-down) ground state TMSM excitations and $n_{\alpha(\beta)}^e$ the number of spin-up (spin-down) excited state TMSM excitations. From this, one can see that the system ground state is given by the ground state TMSM vacuum $\ket{0,0}_\mathrm{sq}^g$. Finally, we remark that the stability of the ground state and excited state TMSM requires $2\left|C\right|\leq\min\left[A+B-\left(\chi_{a}-\chi_{b}\right),A+B+\left(\chi_{a}-\chi_{b}\right)\right]$ and $\omega_{\alpha}^{g/e}$ and $\omega_{\beta}^{g/e}$ have to remain positive. As a consequence the Bogoliubov coefficients $u_{g/e}$ [Eqs.~\eqref{eq:u_g} and \eqref{eq:u_e}] are real and positive, such that $r_{g/e}$ [Eq.~\eqref{eq:r_g_e}] are real and positive as well.
	
	\subsection{Quantum state-dependent excitation frequencies \label{sec:excitation_frequencies}}	
	\begin{figure}[tb]
		\includegraphics[width=\columnwidth]{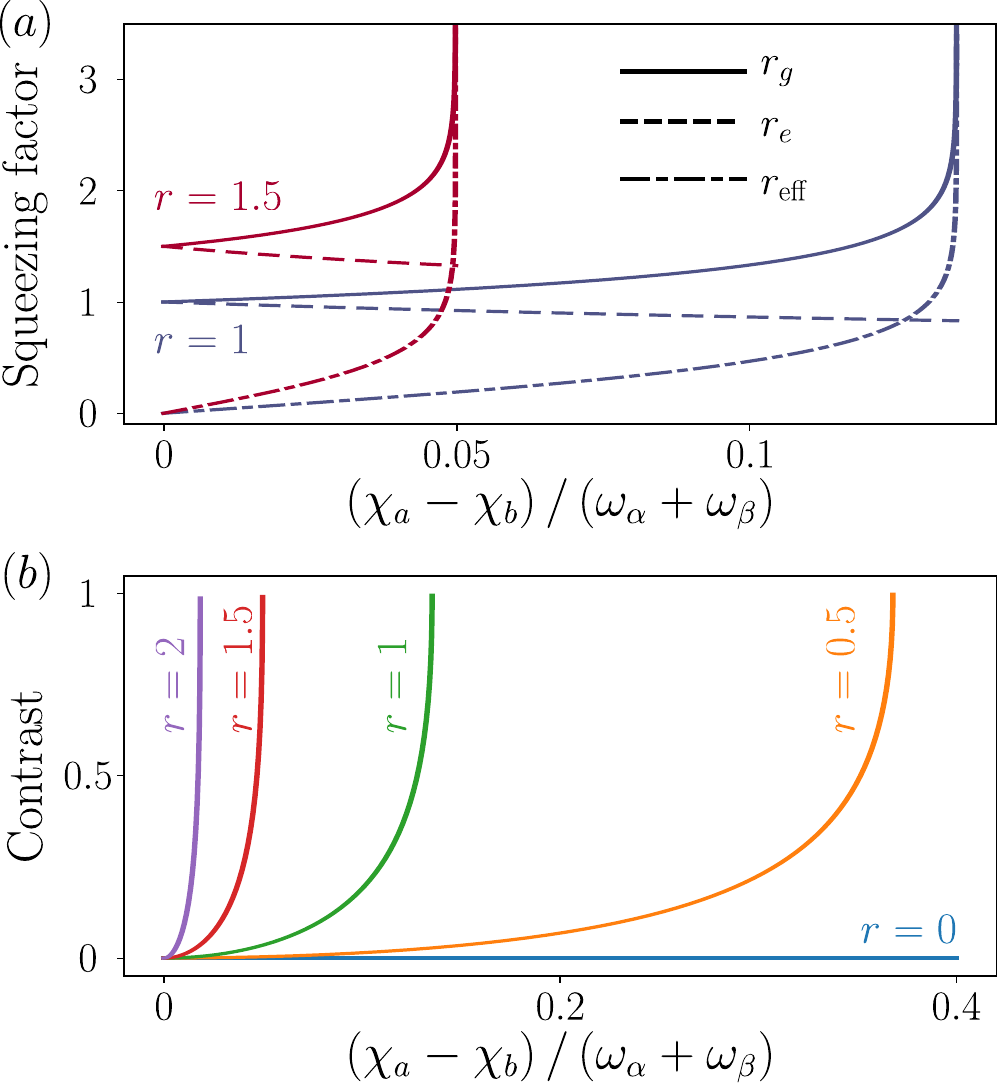}
		\caption{Plot of (a) factors $r_g$, $r_e$ and $r_\mathrm{eff}$ and (b) contrast vs. direct dispersive coupling in units of magnon frequency $(\chi_a-\chi_b)/(\omega_{\alpha} + \omega_{\beta})$. (a) Ground state squeezing $r_g$ (solid line), excited state squeezing $r_e$ (dashed) and the difference $r_\mathrm{eff} = r_g - r_e$ (dashed-dotted) [Eq.~\eqref{eq:r_g_e}] for a bare magnon squeezing of $r=1$ (blue) and $r=1.5$ (red). (b) Contrast $c$ [Eq.~\eqref{eq:contrast}] for several values of $r$.\label{fig:Fig3}}
	\end{figure}
	Our goal is to detect the equilibrium superposition that comprises the nonclassical AFM ground state [Eq.~\eqref{eq:gs_sp}] via qubit spectroscopy as depicted in Fig~\ref{fig:Fig2}(c). In other works such as~\cite{schuster2007,wolski2020,kono2017}, a qubit is coupled dispersively to the eigenmode of a quantum system hosting bosons, like a cavity or a magnet, while a nonequilibrium state is injected into the quantum system. The dispersive coupling induces boson-number-dependent frequency shifts in the qubit, matching the boson number $n$ conserving transitions $\ket{g}\ket{n}\rightarrow\ket{e}\ket{n}$. In the qubit spectrum, the injected state reveals itself as excitation peaks at the shifted qubit frequency representing each eigenmode number state contribution to the superposition comprising the injected state. Here, the idea is that the qubit couples dispersively to noneigenmodes -- the sublattice magnons $\hat{a}$ and $\hat{b}$ -- via $\left(\chi_a\hat{a}^\dagger\hat{a}-\chi_b\hat{b}^\dagger\hat{b}\right)\hat{\sigma}_z$ [Eq.~\eqref{eq:H_0}] in order to resolve the noneigenmode composition of the AFM ground state. As shown for ferromagnets~\cite{romling2023}, the dispersive coupling to the noneigenmode leads to a eigenmode-number-dependent frequency shift in the system resonance frequencies. Qubit excitation $\ket{g}\rightarrow\ket{e}$ is not eigenmode-number-conserving [see Fig.~\ref{fig:Fig2}(b)]. In this procedure, the qubit is driven by a weak monochromatic drive, reading out the qubit excitation in the steady state under the drive. This measurement is performed for a range of frequencies in order to take the qubit spectrum [Fig.~\ref{fig:Fig2}(c)]. If squeezing is present in the magnet, the qubit spectrum contains nontrivial peaks~\cite{romling2023}. In the following, we demonstrate theoretically that this protocol can also be used for AFMs and predict under which conditions nontrivial peaks arise.    
	
	In Sec.~\ref{sec:coupled_system}, we theoretically demonstrate that the ground state of $\hat{\mathcal{H}}_0$ [Eq.~\eqref{eq:H_0}] is the ground state TMSM vacuum $\ket{0,0}_\mathrm{sq}^{g}$. Our goal is to find out into which states the ground state $\ket{0,0}_\mathrm{sq}^{g}$ can be excited by driving the qubit $\ket{g} \rightarrow \ket{e}$ with a monochromatic microwave drive. We model the contribution of the qubit drive to the Hamiltonian via $\hat{\mathcal{H}}_d = \Omega_d \cos(\omega_d)\left(\hat{\sigma}_+ + \hat{\sigma}_- \right)$ where $\Omega_d$ denotes the drive amplitude or Rabi frequency and $\omega_d$ the drive frequency. The Hamiltonian describing the full driven system now reads $\hat{\mathcal{H}}_{\mathrm{full}} = \hat{\mathcal{H}}_0 + \hat{\mathcal{H}}_d$.
	
	As discussed in Sec.~\ref{sec:coupled_system}, there are two types of two-mode squeezing if direct dispersive coupling strengths have different values $\chi_a\neq\chi_b$. If the qubit is in the ground state $\ket{g}(\ket{e})$ the two-mode squeezing is $\xi_g(\xi_e)$ [see Eq.~\eqref{eq:r_g_e} and the schematic in Fig.~\ref{fig:Fig2}(b)]. As a consequence, there exist two subspaces with a set of Fock states each, namely the ground state (excited sate) TMSM Fock states $\left\{\ket{n_\alpha^{g(e)}, n_\beta^{g(e)}}_\mathrm{sq}^{g(e)}\right\}$. 
	\alr{As demonstrated in the appendix,} the vacuum states $\ket{0, 0}_\mathrm{sq}^{g}$ and $\ket{0, 0}_\mathrm{sq}^{e}$ are connected by a two-mode squeeze operation [Eq.~\eqref{eq:S2}] via $\ket{0, 0}_\mathrm{sq}^{g} = \hat{S}_2 \left( \xi_\text{eff}\right) \ket{0, 0}_\mathrm{sq}^{e}$. Here, the factor $\xi_\mathrm{eff}$ denotes the relative squeezing between the vacuum states $\ket{0, 0}_\mathrm{sq}^{g}$ and $\ket{0, 0}_\mathrm{sq}^{e}$. We find that it reads $\xi_\mathrm{eff} = r_\mathrm{eff} e^{i\phi}$ with the absolute value given by $r_\text{eff}=r_g-r_e$. Choosing the convention $\chi_a>\chi_b$ results in $r_g>r_e$, such that the difference between the absolute values of the squeeze factors is positive $r_\text{eff}>0$. For small $\left|\chi_a - \chi_b\right|\ll \min[\left|2\varepsilon^2 /(A+B)\right|, \left|\alr{4}\varepsilon^2 /(A+B+2\varepsilon)\right|]$ this parameter becomes approximately 
	\begin{equation}
		r_\text{eff} = \alr{\sinh(2r)}\frac{\chi_a - \chi_b}{\omega_\alpha+\omega_\beta} , \label{eq:r_eff}
	\end{equation}
	and is an odd function of $\chi_a - \chi_b$. In Fig.~\ref{fig:Fig3}(a) we plot the full expression of $r_\text{eff} = r_g - r_e$ [Eq.~\ref{eq:r_g_e}] as a function of $\left(\chi_a - \chi_b\right)/\left(\omega_\alpha+\omega_\beta\right)$ comparing it with $r_g$ and $r_e$. 
	
	Here we note, that the relatively simple relation $\ket{0, 0}_\mathrm{sq}^{g} = \hat{S}_2 \left( \xi_\text{eff}\right) \ket{0, 0}_\mathrm{sq}^{e}$ is a result of the fact that the phase $\phi$ is the same for $\xi_g$ and $\xi_e$ which simplifies the Baker-Campbell-Hausdorff formula~\cite{achilles2012,romling2023}. In Sec.~\ref{sec:uncoupled_afm}, we discuss that the vacuum state of the uncoupled AFM can be expanded in the sublattice-magnon basis~\cite{kamra2019,walls2008,gerry2005}, as presented in Eq.~\eqref{eq:gs_sp}. We find that a similar expansion of the ground state TMSM vacuum $\ket{0,0}_\mathrm{sq}^g$ is possible in terms of excited state TMSM Fock states $\ket{n_\alpha^e, n_\beta^e}_\mathrm{sq}^e$, following from the relation $\ket{0, 0}_\mathrm{sq}^{g} = \hat{S}_2 \left( \xi_\text{eff}\right) \ket{0, 0}_\mathrm{sq}^{e}$. We obtain the following expression 
	\begin{equation}
		\ket{0,0}_\mathrm{sq}^{g}	=\sum_{n} c_n \ket{n,n}_\mathrm{sq}^{e}, \label{eq:superposition}
	\end{equation}
	with the factors
	\begin{equation}
		c_n	=\frac{\left[-e^{i\phi}\tanh(r_{\text{eff}})\right]^{n}}{\cosh(r_{\text{eff}})}. \label{eq:c_n}
	\end{equation}
	If $r_\text{eff} = 0$ then $c_n=0$ for $n\geq1$, trivializing the superposition [Eq.~\eqref{eq:superposition}] into $\ket{0, 0}_\mathrm{sq}^{g}=\ket{0, 0}_\mathrm{sq}^{e}$. In conclusion, a nontrivial superposition will only occur if $\chi_a \neq \chi_b$ is fulfilled. 
	
	As one can see from Eq.~\eqref{eq:superposition}, there is a nonzero overlap between the system ground state $\ket{0,0}_\mathrm{sq}^g$ and the excited state TMSM Fock states with an equal number of spin-up and spin-down magnons $\ket{n,n}_\mathrm{sq}^e$, given by $c_n = \vphantom{}_\mathrm{sq}^{~e}\!\braket{n,n}{0,0}_\mathrm{sq}^{g}$ [Eq.~\eqref{eq:c_n}]. This allows transitions $\ket{0,0}_\mathrm{sq}^g \rightarrow \ket{n,n}_\mathrm{sq}^e$ from the ground state into an excited state, if the qubit is driven $\ket{g}\rightarrow\ket{e}$ and the drive frequency $\omega_d$ is matching the energy difference between the states $\ket{0,0}_\mathrm{sq}^g$ and $\ket{n,n}_\mathrm{sq}^e$ [see the schematic in Fig.~\ref{fig:Fig2}(b)]. \alr{We denote the energy difference between the states $\ket{0,0}_\mathrm{sq}^g$ and $\ket{n,n}_\mathrm{sq}^e$} as $\omega_n$ \alr{and determine its value via} $\omega_n = \vphantom{}_\mathrm{sq}^{~e}\!\bra{n,n} \hat{\mathcal{H}}_e\ket{n,n}_\mathrm{sq}^{e} - \vphantom{}_\mathrm{sq}^{~g}\!\bra{0,0} \hat{\mathcal{H}}_g\ket{0,0}_\mathrm{sq}^{g}$ with $\hat{\mathcal{H}}_{g/e}$ [Eqs.~\eqref{eq:H_g_diag} and \eqref{eq:H_e}]. \alr{We} obtain the expression
	\begin{equation}
		\omega_{n}	=2n\varepsilon_{e}+\left(\varepsilon_{e}-\varepsilon_{g}\right)-\left(\chi_{a}-\chi_{b}\right)+\omega_{q}\alr{,} \label{eq:omega_n_full}
	\end{equation}
	\alr{where we used $2\varepsilon_{g/e}=\omega_\alpha^{g/e}+\omega_\beta^{g/e}$}. If the resonance condition $\omega_d = \omega_n$ is fulfilled, the transition $\ket{0,0}_\mathrm{sq}^g \rightarrow\ket{n,n}_\mathrm{sq}^e$ occurs with probability $p_n = \left|c_n\right|^2$ [Eq.~\eqref{eq:c_n}], allowing to generate TMSM Fock states $\ket{n,n}_\mathrm{sq}^e$. Since $\tanh(r_\mathrm{eff})<1$, the probabilities $p_n$ are decreasing with increasing $n$. If the qubit is driven continuously at $\omega_d = \omega_n$, the qubit excitation reaches a steady state $\langle \hat{\sigma}_+ \hat{\sigma}_-\rangle_\mathrm{ss}$ which is proportional to $p_n$. Sweeping over the qubit frequency $\omega_d$ therefore results in peaks at $\omega_n$ whose heights are proportional to the corresponding transition probability $p_n$ [see schematic in Fig.~\eqref{fig:Fig2}(c)]. We refer to the described procedure as qubit spectroscopy~\cite{schuster2007,kono2017,romling2023} and conclude that the superposition [Eq.~\eqref{eq:superposition}] reveals itself as nontrivial excitation peaks at transition frequencies $\omega_0,\omega_1,\dots$ [Eq.~\eqref{eq:omega_n_full}]. To quantify the visibility of the first nontrivial peak, we define contrast as $c = \left|c_1\right|^2/\left|c_0\right|^2$ (with factors $c_n$ [Eq.~\eqref{eq:c_n}]) which reads explicitly 
	\begin{equation}
		c	=\tanh(r_{\text{eff}})^{2}. \label{eq:contrast}
	\end{equation} 
	We plot contrast [Eq.~\eqref{eq:contrast}] against $\left(\chi_a - \chi_b\right)/\left(\omega_\alpha+\omega_\beta\right)$ for several values of squeezing $r$ in Fig.~\ref{fig:Fig3}(b). For small $\left|\chi_a - \chi_b\right|\ll \min[\left|2\varepsilon^2 /(A+B)\right|, \left|\alr{4}\varepsilon^2 /(A+B+2\varepsilon)\right|]$, the contrast [Eq.~\eqref{eq:contrast}] can be expanded resulting in 
	\begin{equation}
		c \approx \alr{\sinh(2r)^2}\frac{\left( \chi_a - \chi_b \right)^2}{\left(\omega_\alpha+\omega_\beta\right)^2}. \label{eq:contrast_taylor}
	\end{equation}
	This expression [Eq.~\eqref{eq:contrast_taylor}] shows explicitly the dependence of contrast on $\chi_a-\chi_b$. This confirms that $\chi_a$ and $\chi_b$ have to be finite but different ($\chi_a\neq\chi_b$) in order to observe nontrivial peaks in qubit spectroscopy. While the ratio $(\chi_a - \chi_b )/(\omega_\alpha+\omega_\beta)$ is expected to be small due to large AFM frequencies, the factor $\sinh(2r)^2$ in Eq.~\eqref{eq:contrast_taylor} provides a squeezing-mediated amplification expected to be large due to typically large AFM squeezing~\cite{qin2018,yuan2020}.
	
	Under the condition $\left|\chi_a - \chi_b\right|\ll \left|2\varepsilon^2 /(A+B)\right|$ the expansion of system excitation energies $\omega_n$ [Eq.~\eqref{eq:omega_n_full}] reads
	\begin{align}
		\omega_{n}	\approx n&\left[2\varepsilon+\cosh(2r)\left(\chi_{a}-\chi_{b}\right)\right] \nonumber \\ 
		&+2\sinh^{2}r\left(\chi_{a}-\chi_{b}\right)+\omega_{q}. \label{eq:omega_n_taylor}
	\end{align}
	This expression [Eq.~\eqref{eq:omega_n_taylor}] shows that the separation of two peaks in qubit spectroscopy [Fig.~\ref{fig:Fig2}(c)] is given by the average TMSM frequency $\sim\varepsilon$ of the uncoupled antiferromagnet. 
	
	To conclude this section, we demonstrated that the qubit state can control the magnon squeezing if $\chi_a \neq \chi_b$, leading to nontrivial overlaps between the system ground state $\ket{0,0}_\mathrm{sq}^g$ and excited states $\ket{n,n}_\mathrm{sq}^e$. Driving the qubit at frequencies $\omega_d = \omega_n$ [Eq.~\eqref{eq:omega_n_full}] enables the deterministic generation of TMSM Fock states $\ket{n,n}_\mathrm{sq}^{e}$, where a selection rule imposes that only states with an equal number of spin-up and spin-down magnons can be generated from the ground state. Finally, we demonstrated that nontrivial excitation peaks emerge in qubit spectroscopy if $\chi_a \neq \chi_b$ enabling to resolve the nonclassical AFM ground state.
	
	\section{Physical realizations \label{sec:physical_realizations}}
	In this section, we start by deriving the AFM Hamiltonian $\hat{\mathcal{H}}_\mathrm{AFM}$ [Eq.~\eqref{eq:H_afm}] from a spin model, taking hematite (\alr{$\alpha$-$\mathrm{Fe}_2\mathrm{O}_3$}) as an experimentally available example. We then continue to derive the dispersive coupling [see Eq.~\eqref{eq:H_0}] from an interfacial spin exchange interaction and discuss the role of the interface. 
	\subsection{Hematite \label{sec:hematite}}
	
	Hematite is antiferromagnetic insulator. Its magnetic properties stem from the iron ions while the oxygen ions mediate the exchange between the iron ions~\cite{dannegger2023}. 
	Below the Morin temperature $T_M\approx 250~\text{K}$ hematite crystallizes in the easy-axis antiferromagnetic phase~\cite{morin1950} which is also referred to as rhombohedral phase. 
	
	We consider the spin model with Heisenberg exchange interaction, Dzyaloshinskii-Moriya interaction (DMI), easy-axis single ion anisotropy in second and fourth order and an external magnetic field along the easy-axis. For simplicity, we only take into account the dominating terms in the isotropic Heisenberg exchange interaction. Choosing the $z$-axis as the easy axis, the spin Hamiltonian reads ($\hbar=1$)~\cite{dannegger2023,mazurenko2005,kittel2015,ross2022}
	\begin{align}
		\mathcal{H}_{\mathrm{hem}}	&= \sum_{ l\neq m } \left[ J_{lm}\hat{\boldsymbol{S}}_{l}\cdot\hat{\boldsymbol{S}}_{m} + \mathcal{D}_{lm}\hat{\boldsymbol{z}}\cdot\left(\hat{\boldsymbol{S}}_{l}\times\hat{\boldsymbol{S}}_{m}\right)\right] \nonumber \\
		&- \sum_{l} \left[ K_2\left(\hat{S}_{l}^{z}\right)^{2} + K_4\left(\hat{S}_{l}^{z}\right)^{4}  + \gamma\mu_{0}H_{0}  \hat{S}_{l}^{z}\right], \label{eq:H_hematite}
	\end{align}
	where $\hat{\boldsymbol{S}}_l$ denotes the spin operator at site $l$ in the magnetic lattice, $S$ the total spin on one lattice site, $J_{lm}$ the Heisenberg exchange integral and $\mathcal{D}_{lm}$ the DMI coupling strength between lattice sites $l$ and $m$, $K_2$ and $K_4$ denote the second and fourth order single-ion anisotropy constants, $H_0$ is a constant magnetic field in the $z$-direction and $\gamma<0$ denotes the gyromagnetic ratio. 
	
	We transform the spin Hamiltonian $\hat{\mathcal{H}}_\mathrm{hem}$ [Eq.~\eqref{eq:H_hematite}] into a bosonic form by using linearized Holstein-Primakoff transformations~\cite{holstein1940, kittel2015}. We also switch into momentum space via Fourier transformation. Considering lattice sites $i\in$A and $j\in$B, the \alr{linearized} spin operators read
	\begin{align}
		\hat{S}_{i}^{+}	&=\sqrt{\frac{2S}{N}}\sum_{\boldsymbol{k}}e^{-i\boldsymbol{k}\cdot\boldsymbol{r}_i}\hat{a}_{\boldsymbol{k}}^{\dagger},\label{eq:HP1}\\
		\hat{S}_{i}^{-} 	&=\sqrt{\frac{2S}{N}}\sum_{\boldsymbol{k}}e^{i\boldsymbol{k}\cdot\boldsymbol{r}_i}\hat{a}_{\boldsymbol{k}},\label{eq:HP2}\\
		\hat{S}_{i}^{z}	&=-S+\frac{1}{N}\sum_{\boldsymbol{k},\boldsymbol{k}^{\prime}}e^{-i\left(\boldsymbol{k}-\boldsymbol{k}^{\prime}\right)\cdot\boldsymbol{r}_i}\hat{a}_{\boldsymbol{k}}^{\dagger}\hat{a}_{\boldsymbol{k}^{\prime}},\label{eq:HP3}\\
		\hat{S}_{j}^{+}	&=\sqrt{\frac{2S}{N}}\sum_{\boldsymbol{k}}e^{i\boldsymbol{k}\cdot\boldsymbol{r}_j}\hat{b}_{\boldsymbol{k}},\label{eq:HP4}\\
		\hat{S}_{j}^{-}	&=\sqrt{\frac{2S}{N}}\sum_{\boldsymbol{k}}e^{-i\boldsymbol{k}\cdot\boldsymbol{r}_j}\hat{b}_{\boldsymbol{k}}^{\dagger},\label{eq:HP5}\\
		\hat{S}_{j}^{z}	&=S-\frac{1}{N}\sum_{\boldsymbol{k},\boldsymbol{k}^{\prime}}e^{-i\left(\boldsymbol{k}-\boldsymbol{k}^{\prime}\right)\cdot\boldsymbol{r}_j}\hat{b}_{\boldsymbol{k}}^{\dagger}\hat{b}_{\boldsymbol{k}^{\prime}},\label{eq:HP6}
	\end{align}
	where the spin ladder operators are defined by $\hat{S}_{i}^{\pm}=\hat{S}_{i}^{x}\pm i\hat{S}_{i}^{y}$ and $\hat{S}_{j}^{\pm}=\hat{S}_{j}^{x}\pm i\hat{S}_{j}^{y}$ and $N$ denotes the number of sites in one sublattice. Here we assume that both sublattices have an equal number of sites. Note that $\hat{a}_
	{\boldsymbol{k}}$ and $\hat{b}_{\boldsymbol{k}}$ denote the sublattice-magnon modes (as discussed in Sec.~\ref{sec:uncoupled_afm}) with momentum $\boldsymbol{k}$ and $\boldsymbol{r}_i$ denotes the position of lattice site $i$. Inserting the transformations Eqs.~\eqref{eq:HP1}--\eqref{eq:HP6} into Eq.~\eqref{eq:H_hematite} results in 
	\begin{align}
		\hat{\mathcal{H}}=	\sum_{\boldsymbol{k}}A\hat{a}^{\dagger}_{\boldsymbol{k}}\hat{a}_{\boldsymbol{k}}+B\hat{b}^{\dagger}_{\boldsymbol{k}}\hat{b}_{\boldsymbol{k}}+C_{\boldsymbol{k}}^*\hat{a}_{\boldsymbol{k}}\hat{b}_{-\boldsymbol{k}}+C_{\boldsymbol{k}}\hat{a}^{\dagger}_{\boldsymbol{k}}\hat{b}^{\dagger}_{-\boldsymbol{k}}, \label{eq:H_afmk}
	\end{align}
	with the parameters 
	\begin{align}
		A &= SzJ + Sz^\prime J^\prime + 2SK_2 + 4S^3K_4- \gamma \mu_0 H_0,\label{eq:A}\\
		B &= SzJ + Sz^\prime J^\prime + 2SK_2 + 4S^3K_4+ \gamma \mu_0 H_0,\label{eq:B}\\
		C_{\boldsymbol{k}} &= \gamma_{\boldsymbol{k}} Sz\left(J + i\mathcal{D}\right) + \gamma_{\boldsymbol{k}}^\prime Sz^\prime \left(J^\prime + i\mathcal{D}\right) ,\label{eq:C}
	\end{align}
	and with dominant exchange couplings $J$ and $J^\prime$ from third and fourth nearest neighbors~\cite{mazurenko2005}, the DMI coupling strength $\mathcal{D}$, $z$($z^\prime$) being the number of coupled third(fourth) neighbors and structure factors $\gamma_{\boldsymbol{k}}$($\gamma_{\boldsymbol{k}}^\prime$) from third(fourth) neighbors.  
	
	Here, we consider a small magnet. Due to the confinement and its resulting boundary conditions, the magnon modes exhibit standing waves with discrete spectrum and modes with a finite $\boldsymbol{k}$ are well separated from the $\boldsymbol{k}=\boldsymbol{0}$ by a few GHz~\cite{skogvoll2021}. We will therefore keep only $\boldsymbol{k}=\boldsymbol{0}$ in the sum over $\boldsymbol{k}$ in the transformed Hamiltonian $\hat{\mathcal{H}}$ [Eq.~\eqref{eq:H_afmk}]. \ak{The structure factors thus} simplify to $\gamma_{\boldsymbol{0}}=1$ and $\gamma_{\boldsymbol{0}}^\prime=1$. We define the sublattice-magnon modes as $\hat{a} \equiv \hat{a}_{\boldsymbol{k}=\boldsymbol{0}}$ and $\hat{b} \equiv \hat{b}_{\boldsymbol{k}=\boldsymbol{0}}$. This way, we end up with a Hamiltonian in the form of $\hat{\mathcal{H}}_\text{AFM}$ [Eq.~\eqref{eq:H_afm}]. Using theoretical estimates on the exchange coupling and single ion anisotropies from \citet{mazurenko2005} ($z=3$, $z^\prime=6$, $J=25.2$\,meV, $J^\prime=17.5$\,meV, $S=2$, $K_2=112.8\,\mu$eV, $K_4=1.1\,\mu$eV) and the DMI strength $\mathcal{D}\left(z + z^\prime\right)=\gamma \times 2.2\,$T from \cite{boventer2023}, we estimate that the parameters [Eqs.~\eqref{eq:A}--
	\eqref{eq:C}] approximately read $(A+B)/4\pi=87.46\,$THz and $\left|C\right|/2\pi=87.34$\,THz. These values result in an average resonance frequency of $(\omega_\alpha+\omega_\beta)/4\pi=4.53$\,THz and a squeezing of $r=1.826$ for the uncoupled AFM. 

\ak{Here, we have employed the typical linearized Holstein-Primakoff transformations for obtaining the magnon Hamiltonian. For smaller system sizes, this ignoring of the higher-order terms is harder to justify rigorously. However, it is supported by various experiments finding two-dimensional ordered magnets, for example. Hence, we treat this as an uncontrolled approximation here. Furthermore, it has recently been shown that the primary effect of including the higher order terms is to induce a dephasing of the magnonic states~\cite{yuan2022a,yuan2022b}. We leave a careful consideration of such decoherence effects to a future study.}

	In conclusion, we were able to demonstrate that hematite below the Morin temperature realizes the Hamiltonian $\hat{\mathcal{H}}_\mathrm{AFM}$ [Eq.~\eqref{eq:H_afm}].
	
	\subsection{Role of the interface \label{sec:interface}}
	In this section, we derive the dispersive interaction between the AFM and spin qubit from a spin model. We consider an exchange coupling between the interfacial spins of the AFM and the spin qubit. The Hamiltonian of the interfacial exchange reads~\cite{skogvoll2021} 
	\begin{equation}
		\hat{\mathcal{H}}_{\text{int}}	=J_{\text{int},A}\sum_{l\in A}\hat{\boldsymbol{S}}_{l}\cdot\hat{\boldsymbol{s}}_{l} + J_{\text{int},B}\sum_{m\in B}\hat{\boldsymbol{S}}_{m}\cdot\hat{\boldsymbol{s}}_{m}, \label{eq:H_int}
	\end{equation} 
	where $\hat{\boldsymbol{S}}_{l(m)}$ denotes the spin operator at interfacial lattice site $l(m)$,  $\hat{\boldsymbol{s}}_{l(m)}$ the spin operator of the qubit at interfacial lattice site $l(m)$ and $J_{\mathrm{int},A(B)}$ are the exchange coupling strength~\cite{bender2015, takahashi2010,kamra2016a,zhang2012} to the qubit with interfacial lattice site $l\in \mathrm{A}(m\in \mathrm{B})$.
	
	The product between spin operators can be written as $2\hat{\boldsymbol{S}}_{l}\cdot\hat{\boldsymbol{s}}_{l}=\hat{S}^+_{l}\hat{s}^-_{l} + \hat{S}^-_{l}\hat{s}^+_{l} + 2\hat{S}^z_{l}\hat{s}^z_{l}$. \citet{skogvoll2021} demonstrate that terms $\propto \hat{S}^+_{l}\hat{s}^-_{l}$ and $\propto \hat{S}^-_{l}\hat{s}^+_{l}$ result in a coherent interaction. Here we consider a large detuning between the AFM and qubit such that coherent interaction is suppressed~\cite{gerry2005,romling2023} and will be neglected in the further analysis.   
	
	Using the transformations [Eqs.~\eqref{eq:HP1}--\eqref{eq:HP6}], the terms $\propto \hat{\boldsymbol{S}}^z\cdot\hat{\boldsymbol{s}}^z$ provide two terms up to second order in the sublattice magnon operators $\hat{a}$ and $\hat{b}$. As discussed in \cite{skogvoll2021} and \cite{romling2023}, one of these contributions renormalizes the qubit frequency $\omega_q$, whereas the other is proportional to the sublattice-magnon number operators $\hat{a}^\dagger\hat{a}$ and $\hat{b}^\dagger\hat{b}$. Assuming a constant qubit wave function over the interface, we obtain that the magnon number dependent part leads to the direct dispersive interaction~\cite{skogvoll2021, romling2023} $\hat{\mathcal{H}}_{\text{dis}}=\left(\chi_a\hat{a}^{\dagger}\hat{a}-\chi_b\hat{b}^{\dagger}\hat{b}\right)\hat{\sigma}_{z},$ 
	with the coupling strengths
	\begin{align}
		\chi_{a}	&=\frac{J_{\text{int},A}\left|\psi\right|^{2}}{2N}N_{\text{int},A}, \label{eq:chi_a} \\
		\chi_{b}	&=\frac{J_{\text{int},B}\left|\psi\right|^{2}}{2N}N_{\text{int},B}.\label{eq:chi_b}
	\end{align}
	Here $\left|\psi\right|^2=1/(2N)$ denotes the average qubit wave function over the interface~\cite{skogvoll2021} and $N_{\mathrm{int}, A(B)}$ the number of interfacial lattice sites belonging to sublattice A(B). 
	
	From the expressions for $\chi_a$ and $\chi_b$ [Eqs.~\eqref{eq:chi_a} and \eqref{eq:chi_b}], we deduce that the direct dispersive coupling strength depends on the size of the AFM ($\propto 1/N$) and the structure of the interface. Assuming $J_A=J_B$, for a compensated interface ($N_{\text{int},A} = N_{\text{int},B}$) the direct dispersive coupling strengths are the same $\chi_a=\chi_b$ whereas for uncompensated interfaces ($N_{\text{int},A} \neq N_{\text{int},B}$) the coupling strengths are different $\chi_a\neq\chi_b$. For perfectly uncompensated interfaces, one coupling strength is maximized while the other is zero [see Fig.~\ref{fig:Fig1}(c) for illustration]. 
	
	Semiconductor qubits implement the required exchange interaction [Eq.~\eqref{eq:H_int}]~\cite{chatterjee2021}. We therefore want to estimate $\chi_a$ and $\chi_b$ for an AFM with 100 interfacial sites and 6 monolayers coupled to such a qubit. We use an interfacial coupling strength of $J_A=J_B\approx 10\,\text{meV}$ which has been extracted from spin-pumping experiments~\cite{kajiwara2010,czeschka2011,weiler2013}. For a compensated interface, we calculate a direct dispersive coupling strengths of $\chi_a = \chi_b \approx 2.11\,\text{GHz}$. For a perfectly uncompensated interface with $N_{\text{int},B}=0$ and assuming that the layers are perfectly composed of either sublattice A or B, we find $\chi_a = 4.22\,\text{GHz}$ and $\chi_b = 0$.
	
	In conclusion, the size of the AFM and structure of the interface are crucial factors in determining the direct dispersive coupling strengths $\chi_a$ and $\chi_b$ [Eqs.~\eqref{eq:chi_a} and \eqref{eq:chi_b}]. We also conclude that the protocol to resolve the nonclassical ground state composition of an AFM requires that the interface is uncompensated ($\chi_a\neq\chi_b$). Finally, our estimation resulted in a direct dispersive coupling strengths in the GHz regime. 
	\section{Discussion \label{sec:discussion}}
	Having derived the composition of the ground state $\ket{0,0}_\mathrm{sq}^g$ in terms of excited state TMSM Fock states $\ket{n,n}_\mathrm{sq}^e$ [Eq.~\eqref{eq:superposition}] and the excitation energies $\omega_n$ [Eq.~\eqref{eq:omega_n_full}], we predict that qubit spectroscopy can reveal nontrivial excitation peaks. Now, we want to discuss the resolvability of spectroscopy peaks and requirements for realistic systems. 
	
	As discussed in~\cite{romling2023}, the direct dispersive coupling can be realized with spin qubits, such as semiconducting quantum dots~\cite{chatterjee2021} or NV centers. The latter interact with other spins via dipole-dipole interaction~\cite{casola2018,yan2022}. However, it has been shown in~\cite{romling2023} that the direct dispersive coupling strength stemming from dipole-dipole interaction is expected to be vanishingly small. As a consequence, NV centers are not promising candidates for resolving the nonclassical ground state composition of an AFM. We also want to mention that coherent interaction in the dispersive limit~\cite{gerry2005}, as provided by superconducting qubits~\cite{schuster2007,boissonneault2009,zueco2009}, is not able to resolve nonclassical equilibrium states~\cite{romling2023} and therefore not suitable for resolving the AFM ground state composition. 
	
	In Sec.~\ref{sec:interface}, we estimated that the direct dispersive coupling strengths $\chi_a$ and $\chi_b$ between an AFM and a semiconductor qubit is in the GHz regime. This competes with the resonance frequencies of AFMs up to THz~\cite{fukami2020}. The difference of around 3 orders of magnitude between the coupling strength and the AFM frequency lowers the contrast in qubit spectroscopy. The expectedly low $(\chi_a - \chi_b)/(\omega_\alpha+\omega_\beta)$ can be compensated by large two-mode squeezing $r$ via $\sinh(2r)^2$ [see Eq.~\eqref{eq:contrast_taylor}]. For instance the estimated squeezing factor of $r=1.826$ in hematite results in $\sinh(2r)^2\approx 371.00$. However, this is a given property and depends on the material. \alr{While both squeezing $r$ and magnon frequency $\varepsilon$ depend on the exchange integrals $J$ and $J^\prime$, we find that the ratio $\sinh(2r)/\left(\omega_\alpha + \omega_\beta \right) \approx \left[4\left(2SK_2 + 2S^3K_4\right)\right]^{-1}$ is approximately independent of $J$ and $J^\prime$ when considering $J$ and $J^\prime$ to be the dominant energy scales of the system. \ak{From} Eq.~\ref{eq:contrast_taylor}, we see that \ak{the} contrast is approximately given by $c\approx (\chi_a - \chi_b)^2/\left[16\left(2SK_2 + 2S^3K_4\right)^2\right]$. Since the anisotropy is typically in the GHz range, this further suggests that \ak{a large enough} contrast can be achieved.}
	
	Even for magnets with low squeezing $r$ and low $(\chi_a - \chi_b)/(\omega_\alpha+\omega_\beta)$, there is the possibility of amplifying the peak height of the nontrivial peaks at $\omega_1$, $\omega_2,\dots$ by increasing the drive amplitude $\Omega_d$~\cite{romling2023}. The excitation process into states $\ket{n,n}_\mathrm{sq}^e$ for $n>0$ is suppressed by a factor $\left| c_n\right|^2$ allowing for larger $\Omega_d$ while still being in the linear response regime. Since the peaks at $\omega_n$ and $\omega_{n+1}$ are well separated by the AFM resonance $\sim \varepsilon$ ($\sim$THz), we conclude that indeed large $\Omega_d$ can be chosen without affecting the resolvability with peak overlaps. 
	
	\ak{Our proposed protocol here is complementary to similar qubit-based proposals for detecting the quantum nature of a ferromagnetic ground state~\cite{romling2023} or the magnonic excitation~\cite{skogvoll2021}. There are also recent proposals~\cite{lee2023a} that exploit the magnon squeezing-mediated enhancement of light-matter coupling and superradiant phase transition for detecting the nonclassical properties of these states using a photon cavity. Employing ultrafast laser-induced magnetization dynamics has also been suggested as a probe into the nonclassical magnonic properties~\cite{wuhrer2023,zhao2004,zhao2006}. These approaches complement each other and offer different operation regimes as well as platforms.}

	Lastly we want to discuss the requirement of the uncompensated interface [see Fig.\ref{fig:Fig1}(a)], as concluded from the ground state composition [Eqs.~\eqref{eq:superposition} and \eqref{eq:c_n}]. The coupling strength is maximized by a perfectly uncompensated interface. AFMs, can be grown with uncompensated interfaces~\cite{duo2010}. Some van der Waals materials, such as bilayer $\mathrm{CrI}_3$~\cite{huang2017,liu2023}, consist of ferromagnetic layers that are antiferromagnetically coupled and therefore exhibit a maximized $\chi_a$ while $\chi_b=0$. Also hematite, discussed in Sec.~\ref{sec:hematite} can be grown in layers having uncompensated spins at the surface~\cite{padilha2019}. We conclude that the condition $\chi_a\neq\chi_b$ can be fulfilled by real systems. Since the nontrivial excitation peaks depend on $\chi_a - \chi_b$, qubit spectroscopy provides a measure of how compensated the interface is. This is complementary to already existing techniques of probing spin surface structures, such as spin-polarized scanning tunneling microscopy~\cite{kubetzka2005,schneider2021}.
	\section{Conclusions \label{sec:conclusions}}
	We have shown that the recently suggested protocol for resolving the nonclassical ground state composition of a single-mode squeezed vacuum in a ferromagnet~\cite{romling2023} can be adequately adapted for a similar probing of AFMs exhibiting two-mode squeezing~\cite{kamra2019}. We have shown that the qubit excitation peaks, that serve as the experimental probe of the quantum superposition, are well separated by the typically large AFM resonance frequency $\sim\varepsilon$. Furthermore, hematite below the Morin temperature~\cite{morin1950,mazurenko2005,dannegger2023} realizes the bosonic two-mode squeezing Hamiltonian [Eq.~\eqref{eq:H_afm}]~\cite{walls2008,kamra2019} for the $\boldsymbol{k}=\boldsymbol{0}$ mode. We estimated that the direct dispersive coupling strength is achievable to be in the range of $\sim\,$GHz suggesting small contrast in qubit spectroscopy. However large squeezing of the AFM ground state and optimizing the qubit drive can amplify the qubit excitation peaks making them experimentally detectable. We therefore conclude that this protocol has the potential to resolve the ground state composition of AFMs. 
	
	The direct dispersive coupling to a spin qubit also enables the deterministic generation of non-equilibrium Fock states $\ket{n,n}_\mathrm{sq}^{e}$ by driving the qubit at $\omega_d=\omega_n$ [Eq.~\eqref{eq:omega_n_full}]. The deterministic generation of magnon pairs has the potential to furthermore enable use of the intrinsic entanglement. This could, for instance, be utilized to realize magnonic Hong-Ou-Mandel effect~\cite{kostylev2023,hong1987}. 
	
	\begin{acknowledgments}
		We acknowledge financial support from the Spanish Ministry for Science and Innovation -- AEI Grant CEX2018-000805-M (through the ``Maria de Maeztu'' Programme for Units of Excellence in R\&D) and grant RYC2021-031063-I funded by MCIN/AEI/10.13039/501100011033 and ``European Union Next Generation EU/PRTR''. A. E. R. acknowledges that the project that gave rise to these results received the support of a fellowship from ``la Caixa'' Foundation (ID 100010434) with the fellowship code LCF/BQ/DI22/11940029.
	\end{acknowledgments}
	
	\alr{
	\appendix*
	\section{Derivation of $\xi_\mathrm{eff}$ \label{sec:appendix}}
	In this appendix, we demonstrate the relation $\ket{0, 0}_\mathrm{sq}^{g} = \hat{S}_2 \left( \xi_\text{eff}\right) \ket{0, 0}_\mathrm{sq}^{e}$ with more mathematical detail. The two-mode squeezed-magnon
	vacuum states $\ket{0, 0}_\mathrm{sq}^{g}$ and $\ket{0, 0}_\mathrm{sq}^{e}$ can be written in a common basis as 
	\begin{align}
		\ket{0, 0}_\mathrm{sq}^{g} & =\hat{S}_{2}\!\left(\xi_{g}\right)\ket{0,0}_\mathrm{sub},\label{eq:vac_g}\\
		\ket{0, 0}_\mathrm{sq}^{e} & =\hat{S}_{2}\!\left(\xi_{e}\right)\ket{0,0}_\mathrm{sub}. \label{eq:vac_e}
	\end{align}
	where the squeezing factors $\xi_{g/e}=r_{g/e}e^{i\phi}$ have the same phase given by $\phi$ and \ak{different} absolute values $r_{g/e}$. Inverting the relation Eq.~\eqref{eq:vac_e} results in $\ket{0,0}_\mathrm{sub} = \hat{S}_{2}^\dagger\left(\xi_{e}\right)\ket{0, 0}_\mathrm{sq}^{e}$. Inserting this expression for $\ket{0,0}_\mathrm{sub}$ into Eq.~\eqref{eq:vac_g} leads to the relation 
	\begin{equation}\ket{0, 0}_\mathrm{sq}^{g}=\hat{S}_{2}\!\left(\xi_{g}\right)\hat{S}_{2}^{\dagger}\left(\xi_{e}\right)\ket{0, 0}_\mathrm{sq}^{e}. \label{eq:relation1}
	\end{equation}
	Since $\hat{S}_{2}\!\left(\xi_{e}\right)$ is an exponential operator, the hermitian conjugate can be expressed as  $\hat{S}_{2}^{\dagger}\left(\xi_{e}\right)=\hat{S}_{2}\!\left(-\xi_{e}\right)$. Now we can evaluate the product of operators in Eq.~\eqref{eq:relation1} as $\hat{S}_{2}\!\left(\xi_{g}\right)\hat{S}_{2}^{\dagger}\left(\xi_{e}\right)=e^{\hat{X}}e^{\hat{Y}}$ with the exponents 
	\begin{align}
		\hat{X} & =\xi_{g}^*\hat{a}\hat{b}-\xi_{g}\hat{a}^{\dagger}\hat{b}^{\dagger},\label{eq:exp1}\\
		\hat{Y} & =\xi_{e}\hat{a}^{\dagger}\hat{b}^{\dagger} -\xi_{e}^*\hat{a}\hat{b}.\label{eq:exp2}
	\end{align}
	In order to apply Baker-Campbell-Hausdorff formula~\cite{achilles2012}, we evaluate the commutator $\left[\hat{X},\hat{Y}\right]$. We find 
	\begin{equation}
		\left[\hat{X},\hat{Y}\right]  = \left( \xi_g \xi_e^* - \xi_g^* \xi_e\right) \left[\hat{a}\hat{b},\hat{a}^{\dagger}\hat{b}^{\dagger}\right].\label{eq:commutator}
	\end{equation}
	Since $\xi_g=r_ge^{i\phi}$ and $\xi_e=r_ee^{i\phi}$ have the same phase $\phi$, the coefficient in front of the commutator in Eq.~\eqref{eq:commutator} is equal to zero and Eq.~\eqref{eq:commutator} becomes $\left[\hat{X},\hat{Y}\right]=0$. With the Baker-Campbell-Hausdorff formula, we follow that the operator product from Eq.~\eqref{eq:relation1} can be expressed as $\hat{S}_{2}\!\left(\xi_{g}\right)\hat{S}_{2}^{\dagger}\left(\xi_{e}\right)=e^{\hat{X}+\hat{Y}}$. This explicitly reads
	\begin{align}
		\hat{S}_{2}\!\left(\xi_{g}\right)\hat{S}_{2}^{\dagger}\!\left(\xi_{e}\right) & =\exp(\xi_\mathrm{eff}^*\hat{a}\hat{b}-\xi_\mathrm{eff}\hat{a}^{\dagger}\hat{b}^{\dagger}). \label{eq:relation2}
	\end{align}
	where we defined $\xi_\mathrm{eff}=\xi_g-\xi_e$. We note that $\xi_\mathrm{eff}=\left(r_g-r_e\right)e^{i\phi}$ and define $r_{\text{eff}}=r_{g}-r_{e}$. Finally, we find that Eq.~\eqref{eq:relation2} has the form of a two-mode squeeze operator, such that relation Eq.~\eqref{eq:relation1} can be written as $\hat{S}_{2}\!\left(\xi_{g}\right)\hat{S}_{2}^{\dagger}\left(\xi_{e}\right)=\hat{S}_{2}\!\left(\xi_{\text{eff}}\right)$.
	This leads to 
	\begin{equation}\ket{0, 0}_\mathrm{sq}^{g}=\hat{S}_{2}\!\left(\xi_\mathrm{eff}\right)\ket{0, 0}_\mathrm{sq}^{e}. \label{eq:relation3}
	\end{equation}
	which is the relation \ak{employed} in the main text.
	
}

\end{document}